\documentclass[pdflatex,sn-mathphys-num]{sn-jnl}

\usepackage{graphicx}%
\usepackage{multirow}%
\usepackage{amsmath,amssymb,amsfonts}%
\usepackage{amsthm}%
\usepackage{mathrsfs}%
\usepackage[title]{appendix}%
\usepackage{xcolor}%
\usepackage{textcomp}%
\usepackage{manyfoot}%
\usepackage{booktabs}%
\usepackage{algorithm}%
\usepackage{algorithmicx}%
\usepackage{algpseudocode}%
\usepackage{listings}%
\usepackage{subcaption}
\usepackage{booktabs}
\usepackage{comment}

\theoremstyle{thmstyleone}%
\theoremstyle{thmstyletwo}%

\theoremstyle{thmstylethree}%

\begin{document}

\title[Statistical Learning of Pediatric MHED]{Statistical Learning of Pediatric Mental Health–Related Emergency Department Visits Across COVID-19 Pandemic Periods}


\author[1,2]{\fnm{Anqi A.} \sur{Chen}}\email{anqi6@ualberta.ca}

\author[1]{\fnm{X. Joan} \sur{Hu}}\email{joanh@stat.sfu.ca}

\author*[2]{\fnm{Rhonda J.} \sur{Rosychuk}}\email{rhonda.rosychuk@ualberta.ca}

\affil[1]{\normalsize\orgdiv{Department of Statistics and Actuarial Science}, \orgname{Simon Fraser University}, \orgaddress{
\city{Burnaby}, 
\state{British Columbia}, \country{Canada}}}

\affil*[2]{\normalsize\orgdiv{Department of Pediatrics}, \orgname{University of Alberta}, \orgaddress{
\city{Edmonton}, 
\state{Alberta}, \country{Canada}}}


\abstract{This article presents a statistical learning framework for studying the evolution of pediatric mental health–related emergency department (MHED) visit patterns across the pre-, during-, and post-COVID-19 pandemic periods using population-based administrative health records. The MHED records are formulated as zero-truncated recurrent event data, partitioned into three successive time periods. We develop the modeling framework in a stepwise manner, guided by model fit using a collection of MHED records. The resulting framework progresses from nonparametric marginal rate models to more structured Cox-type regression models for characterizing visit patterns. We ultimately apply stratified regression analysis to investigate changes in visit frequencies and covariate effects across pandemic periods, accounting for prespecified period cut-off points and coarsened individual follow-up information. The proposed framework is motivated by and illustrated using pediatric MHED data throughout the article, providing a practical approach for analyzing recurrent healthcare utilization data with evolving temporal patterns.}




\maketitle

\section{Introduction}\label{sec:Introduction}
The World Health Organization (WHO) reported that approximately one in seven people worldwide lived with a mental disorder in 2021, and nearly half of all mental disorders first emerge before age $18$ \citep{WHO}. Despite the importance of early intervention, only one in five Canadian children with mental health conditions receives adequate mental health services and treatment \citep{CMHA}. Children and adolescents experiencing acute mental health crises, particularly those with limited access to appropriate care, often seek urgent treatment through emergency departments (EDs) \citep{Newton2012cmaj,Cappelli_Cloutier2019}. 
Over the past two decades, pediatric mental health–related emergency department (MHED) visits have increased substantially \citep{CMHA,increase_visit1,covid-19ED_increase,covid-19ED_increase2}. More recently, the COVID-19 pandemic has further affected the mental and emotional well-being of children and adolescents, leading to additional changes in ED utilization \citep{covid-19ED_increase,covid-19ED_increase2,covid-19ED_increase3}. This study is motivated by potential changes in pediatric MHED visits across the COVID-19 pandemic periods.

We formulate MHED records as recurrent event data and focus on statistical models and methods based on event counts. Broadly speaking, there are two major classes: marginal rate/ mean models and intensity-based models. The key distinction is that intensity-based models depend on a subject's event history, whereas marginal models do not; a notable exception is the Poisson process, for which the marginal rate function and the intensity function are identical \cite{Cook.Lawless.2007}. The \textit{Andersen-Gill (AG) model} \citep{AGmodel} extends the Cox proportional hazards model \citep{Cox_reg_model} to recurrent event data through an intensity-based framework. The \textit{Prentice-Williams-Peterson (PWP) models} \cite{PWP1981} are two classes of stratified Cox regression models, using either total follow-up time or gap time to define the baseline time scale. These Cox-type models have been applied to studies of ED utilization \citep{EDvisit_AGmodel,EDvisit_AGmodel2,EDvisit_PWP,EDvisit_PWP_more}. 

In the context of pediatric MHED visits, Hu and Rosychuk \cite{Hu_Rosychuk2016} proposed a Cox-type model with time-varying regression coefficients for analyzing \textit{doubly censored} (both left- and right-censored) recurrent event data with coarsened censoring information resulting from unavailable birthdates, and developed a local linear estimation procedure for estimating the time-varying coefficients. Xiong, Hu, and Rosychuk  \cite{Yi_Hu_Rosychuk2020} further extended this estimation procedure to both doubly censored and zero-truncated recurrent event data, incorporating population census information to account for subjects without MHED visits. Here, \textit{zero truncation} means that a study subject and their observation period are observed only if the subject experiences at least one event \cite{Hu&Lawless1966_supinf_rateMean}. Chen, Hu, and Rosychuk \cite{AC_paper1_suppl} and Chen, Hu, Rosychuk, and Zeng \cite{AC_paper2_suppl} proposed stratified Cox-type models with time-constant and time-varying regression coefficients, respectively, by adapting the estimation method of Xiong et al. and integrating zero-truncated recurrent event data with population census information. As part of this study, we apply the approaches proposed by Chen et al. \cite{AC_paper2_suppl} to MHED data partitioned into the pre-, during-, and post-COVID-19 pandemic periods to investigate changes in pediatric MHED visit patterns over time.

This study is motivated by a Pediatric Mental Health Care (PMHC) program, which uses population-based administrative health databases. These databases are not designed for research purposes and have several inherent limitations. For example, they do not contain data on subjects who do not receive healthcare services, and certain personal information for healthcare users is masked because of privacy protections, such as exact birthdates and postal codes. The study datasets include MHED records of Alberta residents under age 18 between April 1, 2010 and March 31, 2025, together with demographic, geographic, and socioeconomic information on the study units. Here, Alberta, Canada's fourth-largest province by both area and population, had a uniform single-payer health system funded by the Alberta Health Care Insurance Plan (AHCIP) throughout the study period. We define three pandemic periods using March 11, 2020 (the date on which the WHO declared COVID-19 a pandemic) and February 14, 2022 (the date on which Alberta lifted school mask requirements) as the two cut-off points. The study goal is to use the available data to investigate pediatric MHED visit patterns among Alberta residents under 18 years of age across the pre-, during-, and post-COVID-19 pandemic periods.

This study makes two principal contributions. Methodologically, we develop a data-driven, stepwise modeling framework for analyzing recurrent pediatric MHED visits by progressively refining the model from a nonparametric approach to a Cox-type model stratified by pandemic period and event history. Stratification by pandemic period enables the investigation of changes in visit frequencies and covariate effects across the pre-, during-, and post-COVID-19 pandemic periods, whereas stratification by event history captures differences between individuals with different visit histories. Scientifically, the proposed analyses provide insights into how visit frequencies and covariate effects evolved across the three COVID-19 pandemic periods and demonstrate how previous MHED visits influenced subsequent visits.

\section{Methods}\label{sec:method}
\subsection{Study Settings and Data}\label{sec:method_data}
The study subjects were Alberta residents under age 18 who had at least one MHED visit between April 1, 2010 and March 31, 2025. The MHED visits were identified using the \textit{International Classification of Diseases}, 10th edition (ICD-10) diagnostic codes \cite{ICD10}. The complete list of ICD-10 codes used for case identification is provided in Table S1 of Supplementary Material A. Throughout the article, these subjects are referred to as the MHED cohort. Our target population consists of all Alberta residents under 18. Since the MHED cohort includes only subjects with at least one visit, it is not a representative sample of the target population. Consequently, supplementary information is required to account for subjects without MHED visits to make valid inferences about the population. 

The data were obtained primarily from two population-based data sources: the National Ambulatory Care Reporting System (NACRS) \cite{NACRS} and the population census datasets. The NACRS dataset contains ED visit records, including demographic information (age and sex), geographic information (e.g., region of residence and urban/rural status), socioeconomic information (deprivation index), and visit characteristics (e.g., visit date and time, disposition, triage level, and diagnosis). Because submission of ED records to NACRS is mandatory in Alberta, we assume that the absence of an MHED visit in the administrative database corresponds to the absence of an MHED visit in practice. Thus, the MHED visit records constitute zero-truncated recurrent event data with respect to the target population. Due to privacy restrictions, exact birthdates were not available in the NACRS dataset; however, integer ages recorded at each visit allow us to infer the interval in which each subject’s birthdate lies. The population census dataset provides aggregated counts of Alberta residents stratified by age, sex, region of residence, urban/rural status, and deprivation index, which can serve as supplementary information for subjects without MHED visits.

In the subsequent analyses, we considered \texttt{sex} (female vs. male), \texttt{region of residence} (Calgary, Edmonton, and the rest of Alberta), \texttt{deprivation status} (less deprived [quintiles 1-3 and unknown] vs. deprived [quintiles 4-5]), and \texttt{urban/ rural status} (rural vs. urban) as covariates. The baseline levels used were female, the rest of Alberta, less deprived, and rural categories were used as the baseline levels. Because these characteristics were relatively stable over the study period, they were treated as time-independent covariates. For subjects in the MHED cohort, covariate values were obtained from their first recorded MHED visit.

\subsection{Statistical Analysis}\label{sec:method_analysis}
\subsubsection{Notation and modeling strategy}\label{sec:method_notation}
Let $\mathcal{P}$ denote the target population, and let $N_i(a)$ represent the cumulative number of MHED visits experienced by subject $i$ up to age $a$, for $i\in \mathcal{P}$ and $0\leq a <18$, with $N_i(0)=0$. Throughout this study, age (measured in years) is used as the time scale. The corresponding counting process is $N(\cdot)=\big\{N(a):a\geq 0\big\}$ with mean function $\mu(a)$ and rate function $\rho(a)$, where $\mu(a)=\int_0^a\rho(u)du$. Let $\mathcal{H}_i(a)=\sigma\big\{N_i(u): 0\leq u<a\big\}$ denote the event history of subject $i$, where $\sigma$ represents a $\sigma$-algebra. We define a history-based stratification variable $S_i(a)=S\big\{\mathcal{H}_i(a)\big\}$, which takes values $s$ in a finite set $\mathcal{S}$ for a fixed $a$; for $a \in [0, 18)$, $S_i(a)$ is a left-continuous and non-decreasing function. In Section \ref{sec:method_history}, we discuss the stratification variable in more detail and illustrate it with an example.  Let $Z_i$ be the vector of covariates, and let $P_i(a)$ indicate the COVID-19 pandemic period, taking values $p=1, 2, 3$, which correspond to the pre\mbox{-}, during\mbox{-}, and post\mbox{-}pandemic periods, respectively. Ignoring the indices for notational simplicity, our statistical goal is to estimate the baseline intensity functions $\lambda_0(a)$ and either the age-constant regression coefficients $\beta$ or the age-varying regression coefficients $\beta(a)$ under the models considered.

\begin{figure}
\centering
\includegraphics[width=1\textwidth]{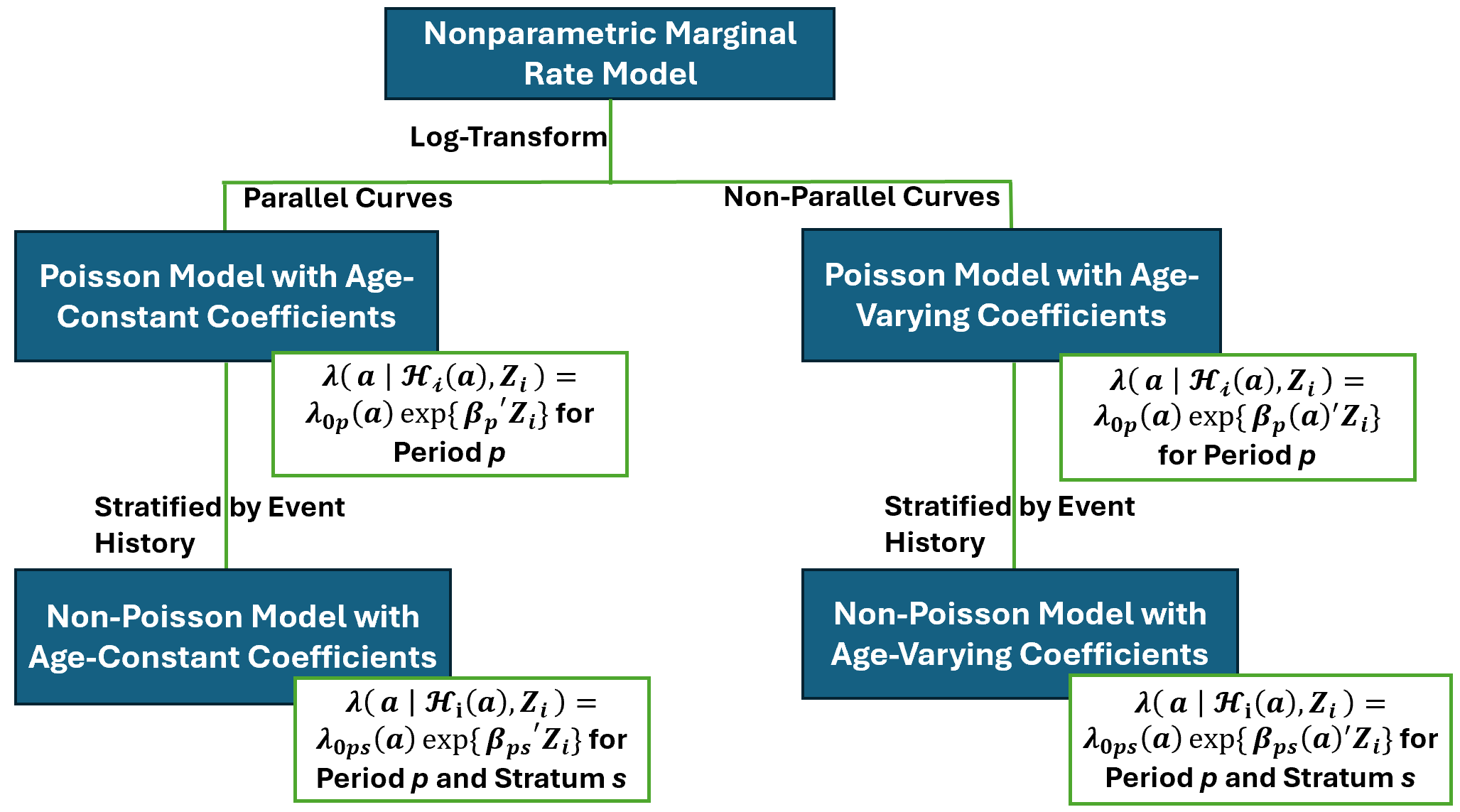}
\caption{\label{Figure:flowchart}Flowchart of the modeling strategy.}
\end{figure}

Fig. \ref{Figure:flowchart} summarizes the proposed stepwise modeling strategy. We begin with minimal modeling assumptions by fitting a nonparametric marginal rate model. The estimated marginal rate functions are then log-transformed to guide subsequent model specification. If the resulting log-rate curves are approximately parallel, we proceed with stratified Cox-type models with age-constant regression coefficients (the left branch of the flowchart). Otherwise, we consider stratified Cox-type models with age-varying regression coefficients (the right branch). The intermediate models are stratified by pandemic period to assess changes in visit frequencies and covariate effects across the pre-, during-, and post-pandemic periods. In the final step, we further stratify the models by event history to investigate whether visit frequencies and covariate effects differ according to a subject's previous MHED visits. In the following subsections, we present the estimation procedures for the nonparametric model and the Cox-type models with age-varying regression coefficients (the root node and the right branch of Fig. \ref{Figure:flowchart}), as the models with age-constant coefficients are special cases of the corresponding models with age-varying coefficients and can be estimated using simpler procedures.

\subsubsection{Nonparametric estimation} \label{sec:method_nonparametric}
We begin with a simple setting in which all required information is fully observed. Under this setting, we derive a nonparametric estimator of a marginal rate function based on a Nelson–Aalen–type estimator of a marginal mean function. We then consider a more realistic setting in which exact birthdates (i.e., the underlying time origin) are unavailable and individual-level covariate information is missing for subjects without MHED visits. In this case, we adapt the rate estimator to accommodate the incomplete information.

Let $\mathcal{O}_1$ denote the MHED cohort and $\mathcal{O}$ be a representative sample of the target population. The population census data provide aggregated covariate information for subjects in $\mathcal{O}$. Let $B_i$ denote the birthdate of subject $i$, and let $[W_{L}, W_{R}]$ represent the data extraction window in calendar time with $W_L=2010-04-01$ and $W_R=2025-03-31$. The corresponding observation window in age for subject $i$ is $(C_{L_{i}}, C_{R_{i}}]$, where $C_{L_{i}}=\max(0, W_L-{B_i})$ and $C_{R_{i}}=\min(18, W_R-{B_i})$. Let $T_{ij}$ denote the calendar time of the $j$th recorded MHED visit of subject $i$. Given the birthdate $B_i$, the event time in age is $A_{ij}=T_{ij}-B_i$. Define  the increment of the counting process $N_i(\cdot)$ by $dN_i(a)=N_i(a)-N_i(a-)$. Then, $dN_i(a)=1$ if subject $i$ experiences an MHED visit at age $a$, and $dN_i(a)=0$ otherwise. With all birthdates available, the observed data can be summarized as $\mathcal{Q}_1=\bigcup_{i\in \mathcal{O}_1}^{} \mathcal{Q}_{1i}=\bigcup_{i\in \mathcal{O}_1}^{}\big[\{dN_i(a):C_{L_i}<a\leq C_{R_i}\}\bigcup_{}^{}\{Z_i\}\big]$. For the remainder of the analysis, we assume that subjects in the target population $\mathcal{P}$ are independent and that both the data extraction window $[W_L,W_R]$ and the birthdate $B_i$ are independent of the counting process $N_i(\cdot)$.

With complete information on birthdates $\boldsymbol{B}=\{B_i:i\in\mathcal{O}\}$ and covariates $\boldsymbol{Z}=\{Z_i:i\in\mathcal{O}\}$, the Nelson-Aalen-type estimator of the marginal mean function $\mu(a)$, for Period $p$ and covariates $z$, is
    \begin{align}
    \begin{split}
    \hat{\mu}_{p}(a,z|\boldsymbol{B},\boldsymbol{Z})&=\int_0^a\frac{\sum_{i\in \mathcal{O}} I(Z_i=z)Y_i^{(p)}(u|{B_i})Y_i^{(c)}(u|{B_i})dN_i(u)}{\sum_{i\in \mathcal{O}} I(Z_i=z)Y_i^{(p)}(u|B_i)Y_i^{(c)}(u|B_i)}\\
    &=\sum_{h:a_{(h)}\leq a}\frac{\sum_{i\in \mathcal{O}_1} I(Z_i=z)Y_i^{(p)}(a_{(h)}|{B_i})Y_i^{(c)}(a_{(h)}|{B_i})dN_i(a_{(h)})}{\sum_{i\in \mathcal{O}} I(Z_i=z)Y_i^{(p)}(a_{(h)}|B_i)Y_i^{(c)}(a_{(h)}|B_i)}, 
    \label{eq:mean_estimator}
    \end{split}
    \end{align}
    where $Y_i^{(c)}(u|B_i)= I\big(u \in (C_{L_i}, C_{R_i}]\big)$, $Y_i^{(p)}(u|B_i)= I\big(u: P_i(u)=p\big)$, and $a_{(1)}<a_{(2)}<\ldots<a_{(H)}$ are the $H$ distinct observed visit times in age. Since $Y_i^{(c)}(\cdot|{B_i})dN_i(\cdot)=0$ for all subjects $i\in\mathcal{O}\backslash\mathcal{O}_1$, the summation over $\mathcal{O}$ can be equivalently restricted to $\mathcal{O}_1$. The corresponding estimator of the marginal rate function $\rho(a)$ is
    \begin{align}
    \hat{\rho}_{p}(a,z|\boldsymbol{B},\boldsymbol{Z})=\frac{\sum_{i\in \mathcal{O}_1} I(Z_i=z)Y_i^{(p)}(a|{B_i})Y_i^{(c)}(a|{B_i})dN_i(a)}{\sum_{i\in \mathcal{O}} I(Z_i=z)Y_i^{(p)}(a|B_i)Y_i^{(c)}(a|B_i)}. 
    \label{eq:rate_idea_estimator}
    \end{align}
At age $a$, the numerator represents the observed number of MHED visits among subjects with covariates $z$ during Period $p$, while the denominator denotes the number of subjects with covariates $z$ who are under observation in Period $p$.

In practice, birthdates $\boldsymbol{B}$ are unavailable in the MHED datasets. We therefore adapt the approach proposed by Hu and Rosychuk \cite{Hu_Rosychuk2016} to handle missing birthdates. Using the recorded integer ages at each MHED visit, $\lfloor A_{ij}\rfloor$, we approximate the birthdate of subject $i\in\mathcal{O}_{1}$ to lie within the interval:
\begin{equation*}
I_i=(W_L-18,W_R]\cap\big\{\cap_{j=1}^{N_i^\star}(T_{ij}-\lfloor A_{ij}\rfloor-1,T_{ij}-\lfloor A_{ij}\rfloor]\big\},
\end{equation*}
where $N^{\star}_i =N_i(C_{R_i})-N_i(C_{L_i})$ denotes the number of observed MHED visits for subject $i$. Following our previous work, we assume that birthdate $B_i$ follows a uniform distribution, $\text{Unif}(I_i)$, for all $i\in \mathcal{O}_1$. We generate $R$ independent birthdates from  $\text{Unif}(I_i)$, denoted by $B_i^{(r)}$ for $r=1,\dots,R$, and substitute them into the corresponding components of Equation (\ref{eq:rate_idea_estimator}). The resulting quantities are then averaged over the $R$ replicates. In addition, we use population census information to approximate the denominator of Equation (\ref{eq:rate_idea_estimator}). The resulting estimator of the marginal rate function for Period $p$ and covariate group $z$ is expressed as: for $a\in(0, 18)$,
       \begin{align}
    \tilde{\rho}_{p}(a,z|\boldsymbol{Z}_1)=\frac{\sum_{i\in \mathcal{O}_1} I(Z_i=z){\frac{1}{R}\sum_{r=1}^R\big\{}Y_i^{(p)}(a|{B_i^{(r)}})Y_i^{(c)}(a|{B_i^{(r)}})dN_i(a)\big\}}{\sum_{l\in\mathcal{L}_p}\mathcal{C}(l,z,\lfloor a \rfloor)},
    \label{eq:rate_estimator}
    \end{align}
where $\boldsymbol{Z}_1=\{Z_i:i\in\mathcal{O}_1\}$ and $\sum_{l\in\mathcal{L}_p}\mathcal{C}(l,z,\lfloor a \rfloor)$ represents the total number of subjects at age $\lfloor a \rfloor$ with covariates $z$ across calendar years $l$ belonging to Period $p$ (i.e., $l\in\mathcal{L}_p$) in the population census data.    

\subsubsection{Estimation under a Poisson model with age-varying regression coefficients}\label{sec:method_period}     
Assume that the conditional intensity function of
the counting process $N_i(\cdot)$ is
\begin{equation}
\lambda(a\mid \mathcal{H}_i(a), Z_i)
=\lambda_{0p}(a) \exp\{\beta_p(a)' Z_i\},
\label{eq:model_period}
\end{equation}
for Period $P_i(a)=p$, $a>0$, and $i\in\mathcal{P}$. Because the intensity function does not depend on the event history $\mathcal{H}_i(a)$, the counting process is a Poisson process. 

To estimate the age-varying regression coefficients $\boldsymbol{\beta}(\cdot)=\{\beta_p(\cdot):p=1,2,3\}$ under Model (\ref{eq:model_period}), we adapt the local linear estimation procedure of Hu and Rosychuk \cite{Hu_Rosychuk2016} and apply a local constant (LC) estimation.  Let $0<\tau_L<\tau_R<18$ denote predetermined constants chosen to avoid a boundary problem in the estimation. The basic idea is that, for a fixed $a\in[\tau_L,\tau_R]$, $\beta_p(u)$ can be locally approximated by its zeroth-order Taylor Expansion, $\beta_p(a)$, for $u$ in the neighborhood of $a$. Specifically, let {$\gamma_p=\beta_p(a)$} and $\boldsymbol{\gamma}=\{\gamma_p:p=1,2,3\}$. Define $K_h(\cdot)=K(\cdot/h)/h$, where $K(\cdot)$ is a kernel function and $h$ is the bandwidth. The kernel function assigns greater weight to observations with age closer to $a$ and zero weight to observations outside the neighborhood of $a$ determined by the bandwidth $h$. The bandwidth controls the smoothness of the estimated coefficient functions, with a larger bandwidth producing smoother estimates.
       
For a fixed $a\in[\tau_L,\tau_R]$ and given birthdates $\boldsymbol{B}_1=\{B_i:i\in\mathcal{O}_1\}$, the estimating function for the regression coefficient $\gamma_p$ is given by
\begin{align}
\begin{split}
U_{p}(\boldsymbol{\gamma};a|\boldsymbol{B}_1)
= \sum_{i\in \mathcal{O}_1}^{}\int_{0}^{18} K_h(u-a) \big\{Z_i
-{\tilde{\bar{Z}}_p(\boldsymbol{\gamma};u)}\big\} Y_i^{(p)}(u|B_i)Y_i^{(c)}(u|B_i) dN_i(u),
    \label{eq:EF_coef_period}
    \end{split}
    \end{align}
where $\tilde{\bar{Z}}_p(\boldsymbol{\gamma};u)=\tilde{G}_p^{(1)}(\boldsymbol{\gamma};u)/\tilde{G}_p^{(0)}(\boldsymbol{\gamma};u)$ with $\tilde{G}_p^{(q)}(\boldsymbol{\gamma};u)=\sum_{z\in \mathcal{Z}}^{}\Bigl\{
z^{\otimes q}\exp\{\gamma_p^{'}z\}$ $\bigl[\sum_{l\in\mathcal{L}_p}\mathcal{C}(l,z,\lfloor u \rfloor)\bigr]\Bigr\}$. Here, $b^{\otimes 0}=1$ and $b^{\otimes 1}=b$. We can solve $U_{p}(\boldsymbol{\gamma};a|\boldsymbol{B}_1)=\mathbf{0}$ for $\gamma_p$ to obtain the estimator $\hat{\beta}_p(a)$ for $p=1,2,3$ and for $a\in[\tau_L,\tau_R]$. We extend the resulting estimator from $[\tau_L,\tau_R]$ to $(0,18)$ by setting $\hat{\beta}_p(a)=\hat{\beta}_p(\tau_L)$ for $a\in(0,\tau_L)$ and $\hat{\beta}_p(a)=\hat{\beta}_p(\tau_R)$ for $a\in(\tau_R,18)$.

Given $\boldsymbol{\beta}(\cdot)$ and birthdates $\boldsymbol{B}_1$, the estimator for the cumulative baseline intensity function is
\begin{align}
\begin{split} 
 \widehat{\Lambda}_{0p}(a\big|\boldsymbol{\beta}(\cdot),\boldsymbol{B}_1)=\sum_{i\in \mathcal{O}_1}\int_{0}^{a} \frac{Y_i^{(p)}(u|B_i)Y_i^{(c)}(u|B_i)}{\tilde{G}_p^{(0)}(\boldsymbol{\beta}(\cdot);u)}dN_i(u),
 \end{split}
 \label{eq:baseline_EF_period}
\end{align}
for $a\in (0,18)$ and $p=1,2,3$. Substituting $\hat{\boldsymbol{\beta}}(\cdot)=\{\hat{\beta}_{p}(\cdot):p=1,2,3\}$ into the expression, we obtain the Breslow-type estimator $\hat{\Lambda}_{0p}(\cdot\big|\hat{\boldsymbol{\beta}}(\cdot))$ \citep{breslow_est1972}. 

When the birthdates $\boldsymbol{B}_1$ are not available, we adapt Approach A proposed by Hu and Rosychuk \cite{Hu_Rosychuk2016} and generate $R$ sets of birthdates $\boldsymbol{B}_1$ from $\text{Unif}(I_i)$ for all $i\in \mathcal{O}_1$, denoted by $\boldsymbol{B}_1^{(r)}$ for $r=1,\ldots,R$. We then can solve $\sum_{r=1}^RU_{p}(\boldsymbol{\gamma};a|\boldsymbol{B}_1^{(r)})/R=\mathbf{0}$ for $\gamma_p$, instead. The remainder of the estimation procedure is similar to that for the case with known birthdates $\boldsymbol{B}_1$.

\subsubsection{Estimation under a non-Poisson model with age-varying regression coefficients} \label{sec:method_history} 
We consider the conditional intensity function of
the counting process $N_i(\cdot)$ given by
\begin{equation}
\lambda(a\mid \mathcal{H}_i(a), Z_i)
=\lambda_{0ps}(a) \exp\{\beta_{ps}(a)' Z_i\},
\label{eq:model_S}
\end{equation}
for $P_i(a)=p$, $S_i(a)=s$, $a>0$, and $i\in\mathcal{P}$. The model depends on the event history $\mathcal{H}_i(a)$ only through the stratification variable $S_i(a)$. A simple example of the stratification variable is given by
\begin{align}
    S_i(a)=\begin{cases}1 & N_i(a-) = 0\\2 &N_i(a-) > 0\end{cases},\text{ for } a\geq0.
    \label{eq:stratification_variable}
\end{align}
This variable indicates that a subject starts in Stratum 1 and transitions to Stratum 2 after the first event. 

Our statistical goal is to estimate the baseline intensity functions $\boldsymbol{\lambda}_{0}^\star(\cdot)=\{\lambda_{0ps}(\cdot):p=1,2,3, s\in \mathcal{S}\}$ and 
the regression coefficients
 $\boldsymbol{\beta}^\star(\cdot)=\{\beta_{ps}(\cdot):p=1,2,3, s\in \mathcal{S}\}$ under Model (\ref{eq:model_S}). For a fixed age $a$, let $\gamma_{ps}=\beta_{ps}(a)$ and $\boldsymbol{\gamma}^\star=\{\gamma_{ps}:p=1,2,3, s\in\mathcal{S}\}$. The estimation procedure follows that in Section \ref{sec:method_period}, with the additional conditional probability $P\bigl(Y_i^{(s)}(u)=1|\mathcal{Q}_{1i},B_i\bigr)$. Since event histories are not fully observed, particularly for subjects born before the start of the data extraction window $W_L$, we do not always know the subjects' strata. Nevertheless, the probability of belonging to a certain stratum is always evaluated conditionally on the birthdate $B_i$ and the available data $\mathcal{Q}_{1i}$.

For a fixed $a\in[\tau_L,\tau_R]$ and $s\in\mathcal{S}$, the estimating function for the regression coefficient $\gamma_{ps}$ is
\begin{align}
\begin{split}
\tilde{U}_{ps}(\boldsymbol{\gamma}^\star;a|\boldsymbol{\lambda}^\star_{0}(\cdot),\boldsymbol{B}_1)
= \sum_{i\in \mathcal{O}_1}^{}\int_{0}^{18} K_h(u-a) &{P\bigl(Y_i^{(s)}(u)=1|\mathcal{Q}_{1i},B_i\bigr)}\big\{Z_i\\
&-{\tilde{\tilde{\bar{Z}}}_{ps}(\boldsymbol{\gamma}^\star;u)}\big\}  Y_i^{(p)}(u|B_i)Y_i^{(c)}(u|B_i) dN_i(u),
    \label{eq:EF_coef_partial_s2}
    \end{split}
    \end{align}
where $Y_i^{(s)}(u)=I\big(u: S_i(u)=s\big)$ and $\tilde{\tilde{\bar{Z}}}_{ps}(\boldsymbol{\gamma}^\star;u)=\tilde{\tilde{G}}_{ps}^{(1)}(\boldsymbol{\gamma}^\star;u)/\tilde{\tilde{G}}_{ps}^{(0)}(\boldsymbol{\gamma}^\star;u)$ with $\tilde{\tilde{G}}_{ps}^{(q)}(\boldsymbol{\gamma}^\star;u)=\sum_{z\in \mathcal{Z}}^{}\Bigl\{P\bigl(Y^{(s)}(u)=1|Z=z\bigr)
z^{\otimes q}\exp\{\gamma_{ps}^{'}z\}$ $\bigl[\sum_{l\in\mathcal{L}_p}\mathcal{C}(l,z,\lfloor u \rfloor)\bigr]\Bigr\}$. 
With fixed $\boldsymbol{\beta}^\star(\cdot)$ and the birthdates $\boldsymbol{B}_1$, the cumulative baseline intensity function is estimated by
\begin{align}
\begin{split} 
 \widetilde{\Lambda}_{0ps}(a\big|\boldsymbol{\beta}^\star(\cdot),\boldsymbol{B}_1)=\sum_{i\in \mathcal{O}_1}\int_{0}^{a} \frac{{P\bigl(Y_i^{(s)}(u)=1|\mathcal{Q}_{1i},B_i\bigr)}}{\tilde{\tilde{G}}_{ps}^{(0)}(\boldsymbol{\beta}^\star(\cdot);u)}Y_i^{(p)}(u|B_i)Y_i^{(c)}(u|B_i)dN_i(u).
 \end{split}
 \label{eq:baseline_EF_partial_S2}
\end{align}
Because the conditional probability $P\bigl(Y_i^{(s)}(u)=1|\mathcal{Q}_{1i}, B_i\bigr)$ potentially contains both the baseline intensity functions $\boldsymbol{\lambda}_{0}^\star(\cdot)$ and the regression coefficients $\boldsymbol{\beta}^\star(\cdot)$, we need to jointly solve $\tilde{U}_{ps}(\boldsymbol{\gamma}^\star;a|\boldsymbol{\lambda}^\star_{0}(\cdot),\boldsymbol{B}_1)=\mathbf{0}$ together with Equation (\ref{eq:baseline_EF_partial_S2}) for $p=1,2,3$ and $s\in\mathcal{S}$. To handle missing birthdates, we adapt Procedure A, following the approach described in Section \ref{sec:method_period}.

The analyses in this section were conducted in R (Version 3.6.1) and the estimation procedures were implemented in \textsf{C++} via the \textsf{R} packages \texttt{Rcpp} \cite{Rccp} and \texttt{RcppArmadillo} \cite{RcppArmadillo}.

\section{Results} \label{sec:results}
This section presents the results of the real data analysis for the models fitted according to the stepwise framework illustrated in Fig. \ref{Figure:flowchart}. As described in Section \ref{sec:method_data}, the potential covariates considered in the analyses were \texttt{sex}, \texttt{region of residence}, \texttt{deprivation status}, and \texttt{urban/rural
status}. Because \texttt{region of residence} and \texttt{urban/rural status} were correlated, they should not be included in the same model. We therefore considered the following covariate combinations in the analyses:
\begin{enumerate}
    \item \texttt{sex} and \texttt{region of residence};
    \item \texttt{sex}, \texttt{region of residence}, and \texttt{deprivation status}; and
    \item  \texttt{sex} and \texttt{urban/rural status}.
\end{enumerate}

Based on our exploration, the common covariate effects exhibited similar patterns across the three model specifications within each type of model. For example, the estimated coefficients for sex were nearly identical across the different Poisson model specifications. Additionally, the estimated effect of the urban indicator was very similar to that of the Calgary indicator within each model type. Therefore, for brevity, we present only the results for Covariate Combination 2 in this article.

\subsection{Patient and Visit Characteristics}\label{sec:Result_characteristics}
\begin{sidewaystable}[ht!]
\centering
\caption{Characteristics of the MHED cohort and visits (2010–2025)}

\scriptsize
\label{Tab:PMHC_summary}
\begin{tabular}{crrrrrrrrr}
\hline\hline
\multirow{2}{*}{} 
&\multirow{2}{*}{Period} 
& \multicolumn{2}{c}{Sex}   
& \multicolumn{3}{c}{Region of residence}  
& \multicolumn{2}{c}{Deprivation} 
& \multirow{2}{*}{Total} \\%
\cmidrule(lr){3-9}%

&  & \multicolumn{1}{c}{Male}   &\multicolumn{1}{c}{Female}     & \multicolumn{1}{c}{Edmonton} & \multicolumn{1}{c}{Calgary} &  \multicolumn{1}{r}{The Rest}  &  \multicolumn{1}{r}{Deprived} & \multicolumn{1}{c}{Less deprived}& \\ \hline

 \multirow{4}{*}{Subjects}    
&Overall 
& 32994   & 49487                   
 & 20333  &  30591   & 31557  
& 38865  & 43616
 & 82481\\ 
 
 &1 
  & 22964     & 31787                
 & 13248  &  20274   & 21229  
 & 26410 & 28341
 & 54751\\
 
 &2 
  & 4849  & 9493                   
 & 3310  &  5749   & 5283 
 & 6384   & 7958
  & 14342\\ 
 
 &3 
  & 7545 & 13282                   
 & 5391  &  7454   & 7982  
 & 9610 & 11217
 & 20827\\ \hline
  
\multirow{4}{*}{Visits}  
&Overall 
 & 56665 &104361                  
 & 39835  & 59070   & 62121 
&77234    &83792  
 & 161026\\ 
 
 &1 
  & 38053 &63246                  
 & 24778  & 37501   & 39020 
&49559   &51740  
 & 101299\\
 
 &2 
& 7148    &16256               
 & 5392  & 9220   & 8792  
  &10527  &12877 
 & 23404\\
 
 &3 
  & 11464 &24859                   
 & 9665  & 12349   & 14309 
  &17148   &19175 
 & 36323\\\hline\hline
\end{tabular}
\end{sidewaystable}

The MHED dataset contained 82,481 subjects who experienced a total of 161,026 MHED visits between fiscal years 2010 and 2025. During the pre-pandemic period, 54,751 subjects accounted for 101,299 visits, corresponding to an average of 1.9 visits per person and 10,129.9 visits per year. During the COVID-19 pandemic, the average number of visits per person decreased to 1.6, whereas the average number of visits per year increased to 11,702, with a total of 14,342 subjects contributing 23,404 visits. In the post-pandemic period, the average number of visits per person increased slightly to 1.7, and the average number of visits per year further increased to 12,107.7, based on 20,827 subjects and 36,323 MHED visits. Generally speaking, a longer observation window could lead to more visits per person, whereas the higher annual number of visits more accurately reflected the increasing trend across the three periods.

Table \ref{Tab:PMHC_summary} summarizes the characteristics of the MHED cohort and the corresponding visits. Overall, the cohort consisted predominantly of females (60.0\%), residents living outside Edmonton and Calgary (38.3\%), and subjects from less deprived communities (52.9\%). During the COVID-19 pandemic period, the proportion of females increased to 66.2\%, and a larger proportion of MHED visits were made by residents of Calgary (40.1\%). The proportion of children and youth from less deprived communities also increased slightly to (55.5\%). However, these characteristics do not reflect those of the Alberta population under 18 years of age. For example, males accounted for 51.1\% of the Alberta pediatric population, whereas females comprised 60.0\% of the MHED cohort. This discrepancy further illustrates that the MHED cohort is not representative of the Alberta population under 18 because it includes only subjects with at least one MHED visit.

\begin{figure}[!ht]
    \centering
    \includegraphics[width=0.97\textwidth]{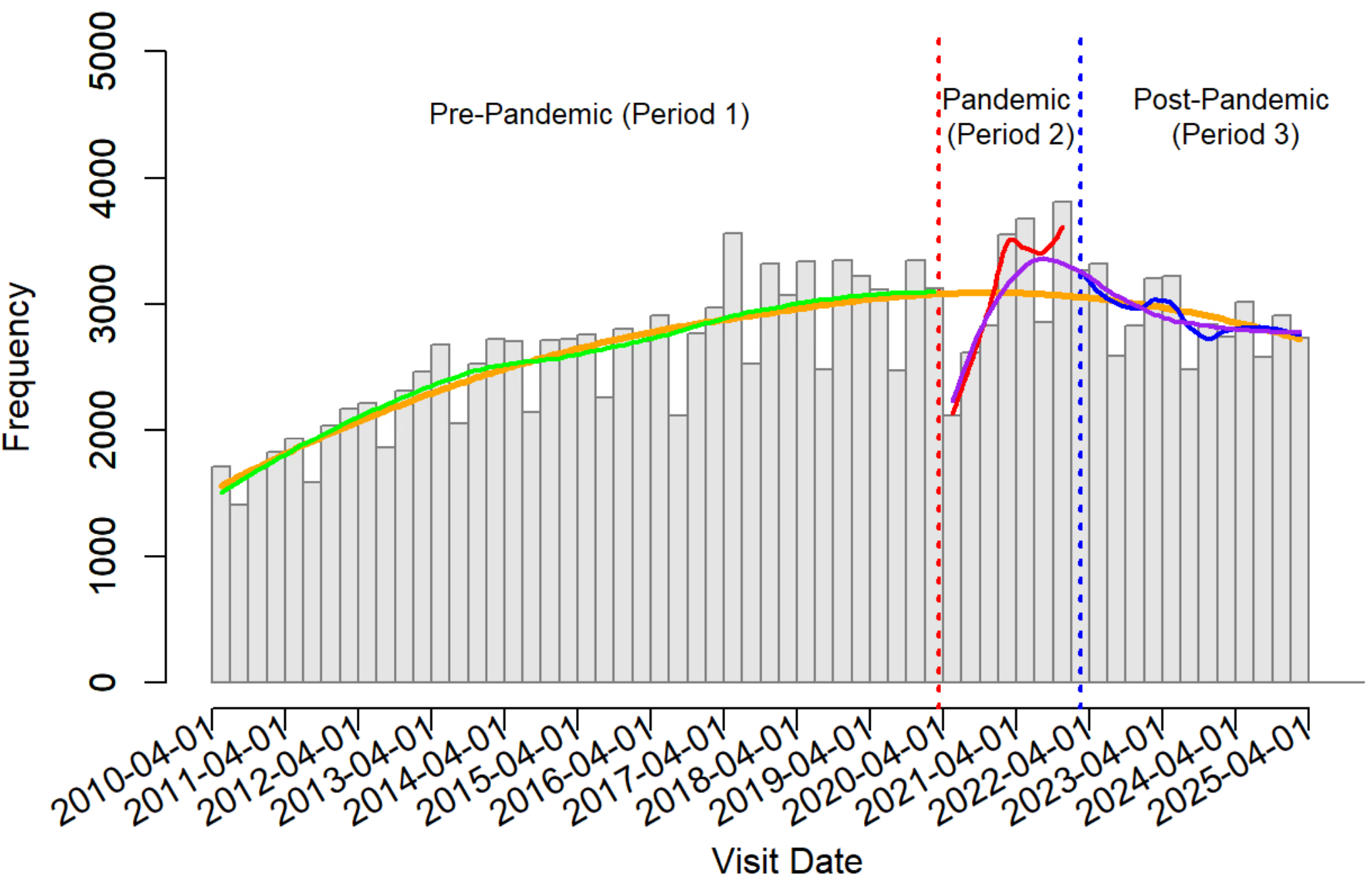}
    \caption{\label{Figure:hist_visit}Histogram of MHED visits over time.}
\vspace{-0.2cm}
\raggedright\scriptsize{\noindent Note: The orange curve represents the trend over the entire observation period. The green, red, and blue curves represent the trends during the pre-, during-, and post-pandemic periods, respectively. The purple curve represents the trend since the onset of the pandemic.}
\end{figure}

The characteristics of MHED visits were generally consistent with those of the MHED cohort (see Table \ref{Tab:PMHC_summary}). Overall, 89.0\% of subjects experienced three or fewer MHED visits during the entire observation period, and 62.6\% had only a single visit. Most MHED visits (82.3\%) occurred among adolescents aged 13 years or older, whereas only 3.3\% were made by children younger than 8 years. Additional information on the age distribution of MHED visits by pandemic period is provided in Fig. S1 of Supplementary Material B. Fig. \ref{Figure:hist_visit} displays the number of MHED visits from fiscal years 2010 to 2025. The orange LOESS (Locally Estimated Scatterplot Smoothing) curve represents the overall trend, whereas the green, red, and blue LOESS curves correspond to the pre-pandemic, pandemic, and post-pandemic periods, respectively. The purple one shows the trend from the onset of the COVID-19 pandemic onward. The number of MHED visits increased gradually during the pre-pandemic period. Following the onset of the COVID-19 pandemic, the number of visits declined abruptly before increasing rapidly. During the post-pandemic period, the number of visits stabilized, with a slight downward trend. A clear seasonal pattern was also observed, with fewer MHED visits occurring in July and August, likely reflecting the summer school break. Consequently, every third bar in the histogram is noticeably shorter.

\subsection{Nonparametric Estimation Results}\label{sec:Result_nonparametric}
Fig. S2 presents the smoothed estimated marginal rate functions by covariate combination and pandemic period, along with the corresponding 95\% pointwise confidence intervals (CIs). The marginal rate represents the instantaneous rate of experiencing an MHED visit at a given age. In the figure, “F” denotes female and “M” denotes male. The abbreviations “TR,” “C,” and “E” represent the rest of Alberta regions, Calgary, and Edmonton, respectively, while “LD” and “D” indicate less deprived and deprived communities. As shown in Fig. S2, subjects living in deprived communities generally had higher estimated marginal rates of MHED visits than those living in less deprived communities, particularly after age 14, across all three pandemic periods. With a fixed deprivation status, females residing outside Calgary and Edmonton had higher estimated marginal rates of MHED visits, whereas males residing in Calgary or Edmonton had lower estimated marginal rates during all three periods.

\begin{figure} 
\begin{subfigure}[t]{0.475\textwidth}
\centering
\includegraphics[width=\linewidth]{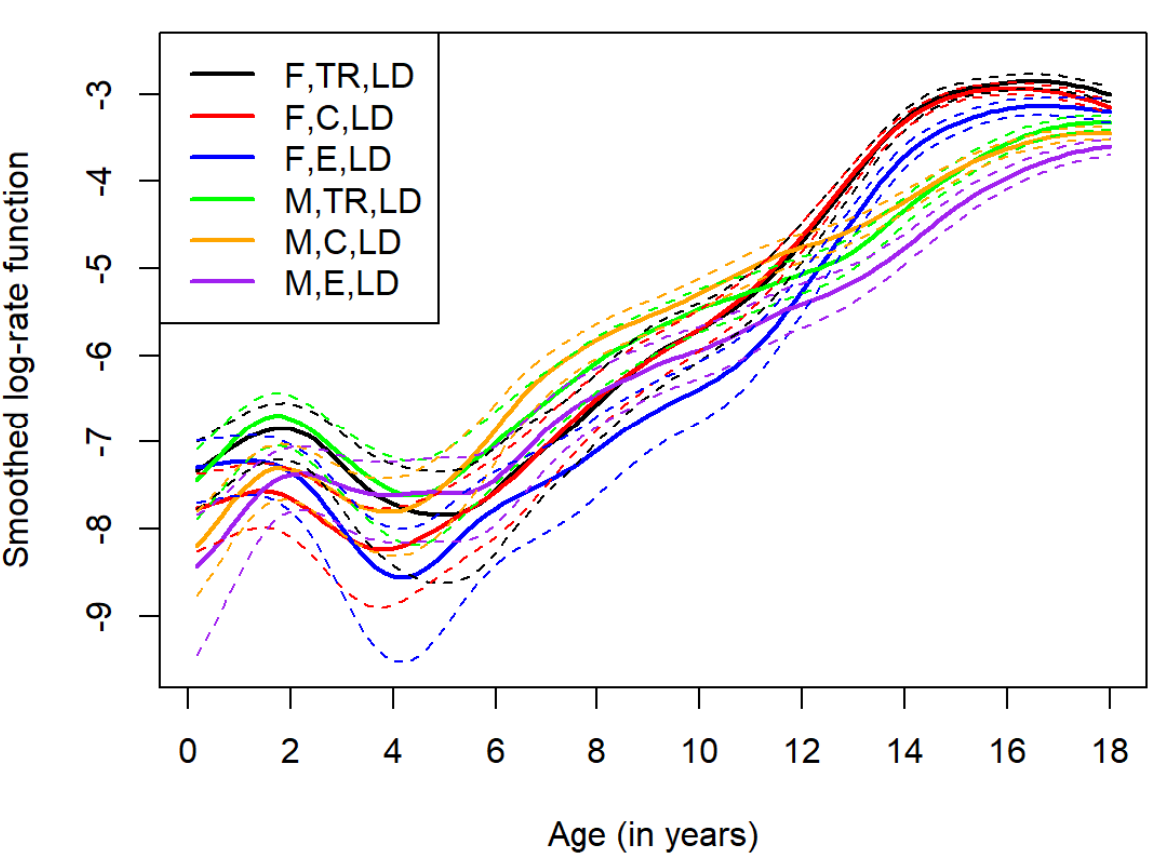}
\caption{\small Less deprived} 
\end{subfigure}\hskip 1em%
\begin{subfigure}[t]{0.48\textwidth}
\centering
\includegraphics[width=\linewidth]{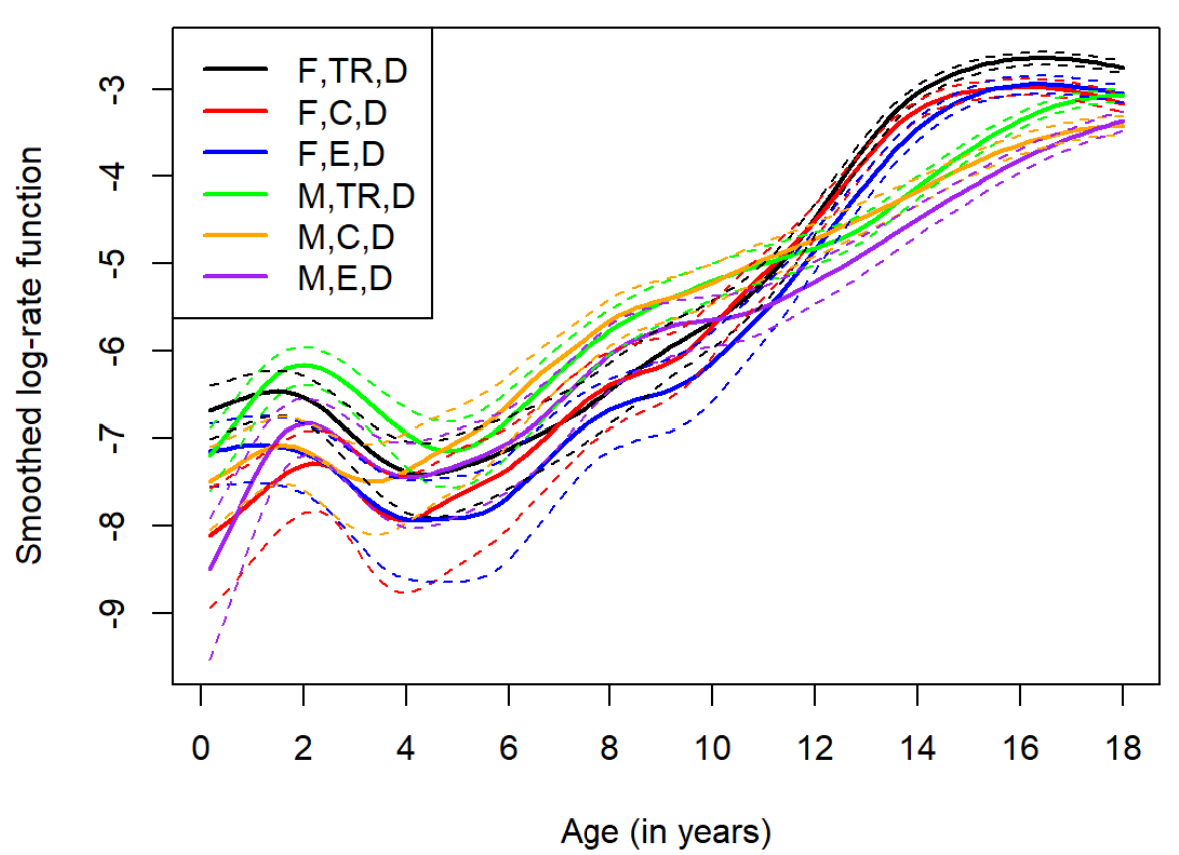}
\caption{\small Deprived} 
\end{subfigure}
\caption{\label{Figure:log_rate_p1} Log-transformed estimated rate functions with 95\% pointwise confidence intervals for Period 1.} 
\vspace{-0.2cm}
\raggedright\scriptsize{\noindent Note: ``F'' and ``M'' represent female and male; ``TR'', ``C'', and ``E'' correspond to the rest of Alberta regions, Calgary, and Edmonton, respectively; ``LD'' and ``D'' are less deprived and deprived communities.}
\end{figure}

Fig. \ref{Figure:log_rate_p1} displays the logarithm-transformed estimated marginal rate functions for the pre-COVID-19 pandemic period; the corresponding results for the pandemic and post-pandemic periods are provided in Fig. S3 of Supplementary Material B. The log-rate curves for females and males crossed each other, indicating that the sex effect varied with age. Therefore, we fitted models with age-varying regression coefficients, as described in Fig. \ref{Figure:flowchart}, in the following subsection.

\subsection{Estimation Results for Models with Age-Varying Regression Coefficients}
Figs. \ref{Figure:coef_fun_sex} and S4 - S5 in Supplementary Material A show the estimated regression coefficients with 95\% pointwise CIs under Models (\ref{eq:model_period}) and (\ref{eq:model_S}). Model (\ref{eq:model_period}), which does not incorporate stratification by event history, corresponds to the left panels of these figures, whereas the middle and right panels correspond to Model (\ref{eq:model_S}). Because most study subjects experienced only one MHED visit, we used the history-based stratification variable defined in (\ref{eq:stratification_variable}). Additionally, we discretized age into two-month units. Set the pre-determined constant $\tau_L$ to 9 units (1.5 years) and the constant $\tau_R$ to 105 units (17.5 years). The Epanechnikov kernel was applied with bandwidth $h=9$ units (equivalent to 1.5 years). In the figures, the dark green, dark red, and black curves represent the estimated age-varying regression coefficients for the pre-, during-, and post-pandemic periods, respectively. For comparison, the corresponding age-constant estimates are also shown. The substantial overlap between the 95\% pointwise CIs for the age-varying and age-constant estimates suggests that an age-constant coefficient may be adequate in some settings. Nevertheless, the results support the use of age-varying regression coefficients. 

\begin{figure}[t!] 
\begin{subfigure}{0.33\textwidth}
\begin{subfigure}{0.99\textwidth}
\includegraphics[width=\linewidth]{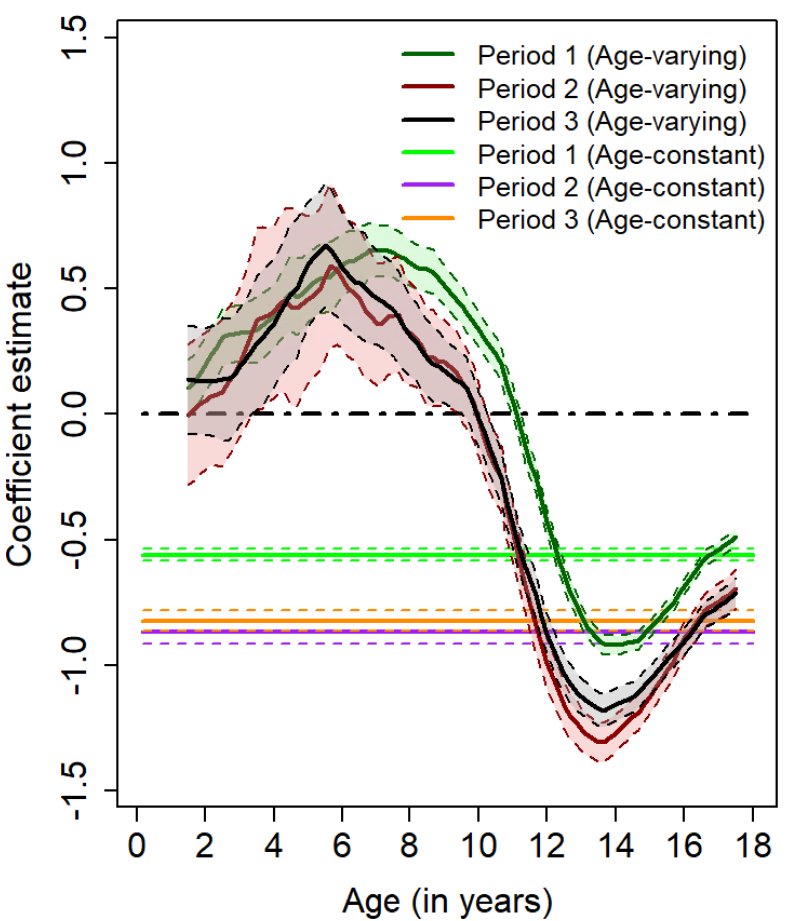}
\end{subfigure}
\vspace{-0.05cm}
\caption{\scriptsize No stratification by history} \label{fig:coef_sex_noS}
\end{subfigure}\hspace*{\fill}
\begin{subfigure}{0.65\textwidth}
\renewcommand{\thesubfigure}{b1}
\begin{subfigure}{0.49\textwidth}
\includegraphics[width=\linewidth]{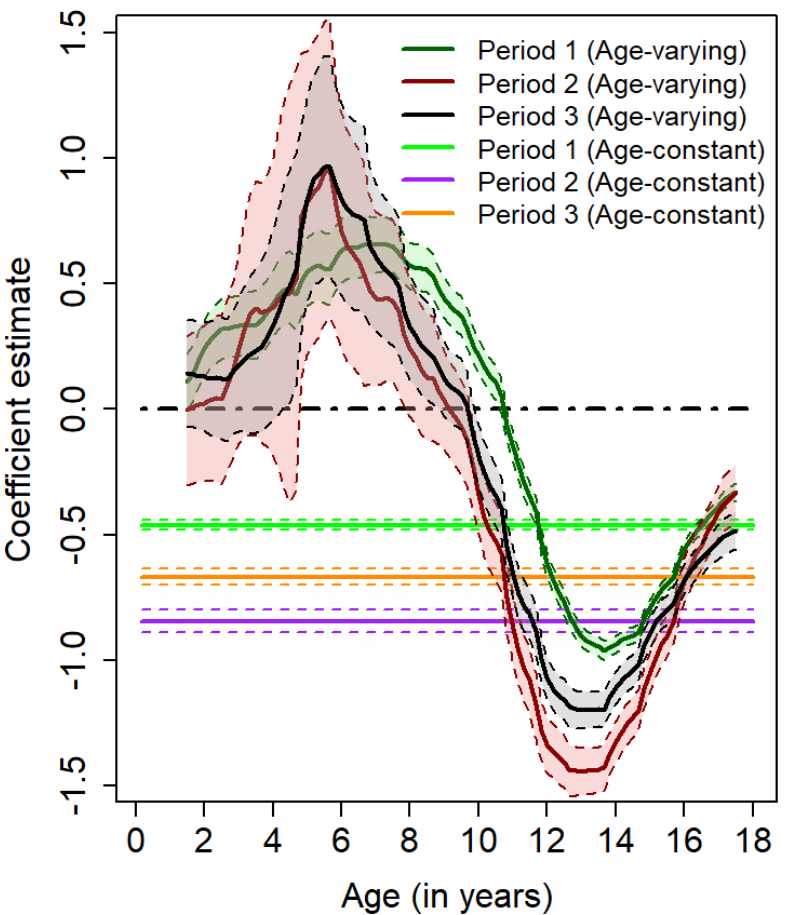}
\caption{\scriptsize Stratum 1} \label{fig:coef_sex_s1}
\end{subfigure}\hspace*{\fill}
\renewcommand{\thesubfigure}{b2}
\begin{subfigure}{0.49\textwidth}
\includegraphics[width=\linewidth]{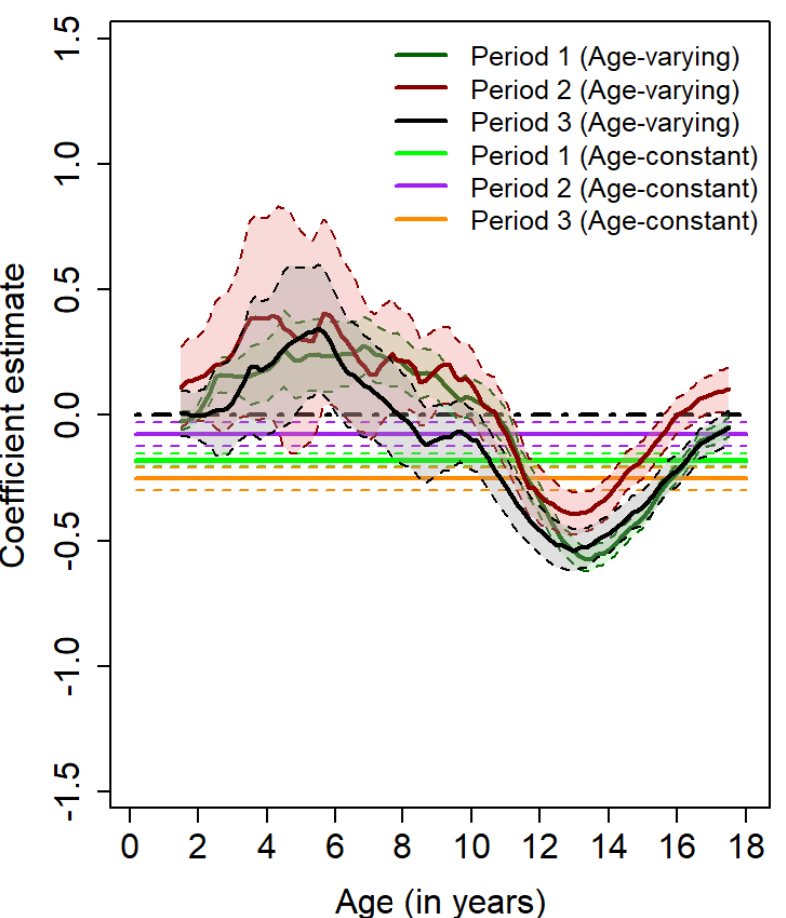}
\caption{\scriptsize Stratum 2} \label{fig:coef_sex_s2}
\end{subfigure}
\vspace{-0.8cm}
\renewcommand{\thesubfigure}{b}
\caption{\scriptsize Stratification by history} \label{fig:coef_sex}
\end{subfigure}
\caption{\label{Figure:coef_fun_sex}Estimated regression coefficients for male (vs. female) with 95\%
pointwise confidence intervals.}
\end{figure}

We first considered the model without history-based stratification. Fig. \ref{fig:coef_sex_noS} shows that, during the pre-pandemic period, boys had a higher risk of MHED visits than girls before approximately 11.2 years of age, whereas girls were at higher risk thereafter, after adjusting for region of residence and deprivation status. Similar patterns were observed during the pandemic and post-pandemic periods, although the turning point switched earlier to about 10 years. Fig. S4a indicates that Edmonton residents generally had lower risks of MHED visits than residents living outside Edmonton and Calgary throughout the three pandemic periods. Fig. S4c shows that Calgary residents had lower risks before age 5 and after age 13 during the pre- and post-pandemic periods, whereas they experienced higher risks between ages 4 and 6 during the pandemic period. Regarding deprivation status, Fig. S5a displays that children from deprived communities generally had higher risks of MHED visits during the pre-pandemic period. During and after the pandemic, however, little difference was observed between deprived and less deprived communities for most ages below 14.  

Under the model with history-based stratification, the interpretation is similar, except that the panels for Stratum 1 correspond to the risk of the first MHED visit, whereas those for Stratum 2 correspond to the risk of subsequent visits under stratification variable (\ref{eq:stratification_variable}). For example, the dark green curve in Fig. \ref{fig:coef_sex_s1} shows that boys were more likely than girls to experience their first MHED visit before approximately 10.8 years of age during the pre-pandemic period, whereas girls were at higher risk thereafter. Fig. \ref{fig:coef_sex_s2} further indicates that, after the first MHED visit, boys had higher risks of subsequent visits before approximately 9.5 years of age but lower risks thereafter than girls. Additional findings include that Calgary residents generally had lower risks of a first MHED visit than residents outside Calgary and Edmonton during the pre- and post-pandemic periods, although children aged approximately 4 to 7.5 years had higher risks during the pandemic (see Fig. S4d1). Furthermore, Calgary residents had higher risks of subsequent MHED visits at younger ages before and during the pandemic but lower risks after the pandemic than residents living outside Calgary and Edmonton (Fig. S4d2). With respect to deprivation status, children younger than 6 years living in deprived communities were at higher risk of experiencing a first MHED visit than those living in less deprived communities only before the COVID-19 pandemic (Fig. S5b1).

The estimated cumulative baseline intensity functions are shown in Fig. S6. The estimated cumulative baseline functions were similar for models differing only in whether the regression coefficients were age-varying or age-constant. Under the history-stratified models, however, the estimated cumulative baseline intensity was substantially higher for Stratum 2 than for Stratum 1. 

The estimated model can also be used to obtain marginal mean functions. As an illustration, Fig. S7 presents the estimated mean function $E[N(a)|Z]$ under Model (\ref{eq:model_period}) for different covariate combinations and pandemic periods. This function represents the expected cumulative number of MHED visits up to a given age. The estimated mean functions exhibited patterns similar to those of the estimated marginal rate functions shown in Fig. S2; therefore, these patterns were not repeatedly discussed here.

\section{Discussion}\label{sec:discussion}

The proposed analyses relied on two main assumptions: (i) independence among subjects in the target population $\mathcal{P}$ and (ii) independent censoring. Regarding the first assumption, one potential concern is that subjects residing in the same community may be correlated. In this study, however, the region of residence was defined using large geographic zones, making such correlations unlikely to substantially affect the results. The second assumption may be more restrictive because the censoring time, determined by the birthdate $B_i$, may not be independent of the counting process $N_i(\cdot)$. For example, Xiong et al. \cite{Yi_Hu_Rosychuk2020} reported generational differences in MHED visit patterns, suggesting that birth cohort may be associated with the event process. A possible extension is to account for informative censoring using inverse probability of censoring weighting (IPCW), which will be investigated in future work.

A potential limitation of this study is that, although the analyses identified changes in visit frequencies and covariate effects across the three pandemic periods, these differences cannot be attributed solely to the COVID-19 pandemic. Instead, they may reflect a combination of pandemic-related effects and underlying temporal trends. Future research is needed to address the potential confounding between trends in MHED visit patterns and the COVID-19 pandemic to better quantify their respective effects.


Several directions for future research warrant further investigation. In the present study, the pandemic periods were defined using prespecified temporal cutoffs. Future work will focus on developing statistical methods that account for uncertainty in these temporal cutoffs while accommodating the coarsened follow-up information available in the administrative databases. In addition, we plan to extend the proposed framework to investigate MHED visits associated with specific mental health conditions, such as intentional self-harm and mood disorders.

\backmatter

\section*{Supplementary information}
Electronic supplementary material is available.

\section*{Acknowledgements}
This study is based, in part, on data provided by Alberta Health. The interpretations and conclusions presented herein are those of the authors and do not necessarily reflect the views of the Government of Alberta or Alberta Health. Neither the Government of Alberta nor Alberta Health endorses or expresses any opinion regarding the findings or conclusions of this study.

\section*{Declarations}
\section*{Funding}
This work was supported by individual Discovery Grants from the Natural Sciences and Engineering Research
Council of Canada (NSERC) held by XJ Hu and RJ Rosychuk.

\subsection*{Competing Interests}
No competing interest is declared.

\subsection*{Ethics Approval}
Not applicable.

\subsection*{Consent to Participate}
Not applicable.

\subsection*{Consent for publication}
Not applicable.

\subsection*{Data Availability}
The data were provided by Alberta Health and are not publicly available. The authors are not permitted to share the data due to privacy and data-sharing restrictions.

\subsection*{Materials Availability}

\subsection*{Code Availability}



\clearpage

\begin{appendices}

\end{appendices}


\clearpage
\bibliography{sn-bibliography}

\end{document}